\UseRawInputEncoding   
 \documentclass[a4paper]{jpconf}
\usepackage{graphicx}
\usepackage{iopams}
\begin{document}
\title{Quantum physics  
	and biology: \\the local wavefunction approach}   

\author{George F R Ellis$^{1,2}$} 

\address{$^1$ Mathematics Department, University of Cape Town, Rondebosch 7700, Cape Town, South Africa.}
\address{$^2$ The New Institute, 18 Warburgstrasse, Hamburg 20354, Germany.}

\ead{george.ellis@uct.ac.za}

\begin{abstract}
Is there a single linearly evolving Wave Function of the Universe that is able to lead to all the nonlinearities we see around us? This proposal seems \textit{a priori} highly implausible. I claim that instead, in the real Universe, generically only local wave functions exist. Non-local wave functions occur for carefully engineered contexts such as Bell experiments, 
 but there is no single wave function for a cat or  macroscopic object such as a brain, let alone for the Universe as a whole. Contextual wave function collapse leads to a defensible version of the Copenhagen interpretation of quantum theory, where classical macro levels  provide the context for quantum events and biological emergence. 
 Complexity arises via multiscale adaptive modular hierarchical structures that enable logical branching to emerge from the underlying linear physics. Each emergent level is causally effective because of the meshing of upwards and downwards causation that takes place consistently with that   physics. 
 Quantum chemistry approaches in biological contexts  fit this local wavefunction picture. 
\end{abstract}

\section{Introduction}

This paper reviews three puzzles to do with the way physics relates to emergent reality. They are  to do with quantum physics, with complexity, and with how these relate to each other.

1. \textbf{The puzzle of quantum linearity 
} How can the highly non-linear complexity we see around us arise  out of a linear equation for a single wave function of the universe, as some claim? I propose that in real world applications only local wave functions exist; there is no single wave function for a living cell, or macroscopic objects such as a cat or brain.
 Contextual wave function collapse leads to a defensible version of the Copenhagen interpretation of quantum theory where classical macro levels  provide the context for quantum events.
This is discussed in Section 2.

2. \textbf{The puzzle of complexity} So how does complexity arise out of local linear dynamics?  This occurs via adaptive modular hierarchical structures, where each emergent level is causally effective because of the meshing of upwards and downwards causation that takes place consistently with the underlying physics via time-dependent constraints. Such emergent structures enable logical branching to emerge from the underlying physics via macromolecular chemistry. 
Causal completeness is not a property of physics \textit{per se}:  
it only occurs through closure of constraints involving all linked emergent levels. This is discussed in Section 3.

3. \textbf{How these relate to each other} Given the previous two understandings, in what way do linear quantum processes underlie biology? They underlie the basic features of the stability of matter, the nature of the periodic table of the elements, chemical bonding, and the folding of biomolecules. Essentially quantum effects occur in a few cases (enzyme action, photon detection, magnetic field detection), but the local wavefunction approach advocated here is applicable in all cases, and leads to the highly non-linear nature of molecular biology that cannot be described by unitary evolution of a single wave function.  This is discussed in Section 4.

4. \textbf{The meta issue} is the domain of application of a theory. This is discussed in Section 5.

\section{The puzzle of quantum linearity}

Quantum theory \cite{Isham} \cite{QM} is essentially a linear (Hamiltonian) theory \cite{Silverman} \cite{Ellis (2012) QM}.
Nowadays it is often claimed (e.g. \cite{Carroll (2022)}) that there is a unitarily evolving wavefunction for the entire universe from which all else follows, 
hence all that happens is based in linear evolution of a single wave function . 

However the real world is not linear either in terms of structures or associated functions (Figure 1).
\begin{figure}[h]
	\centering
	\includegraphics[width=0.6\linewidth]{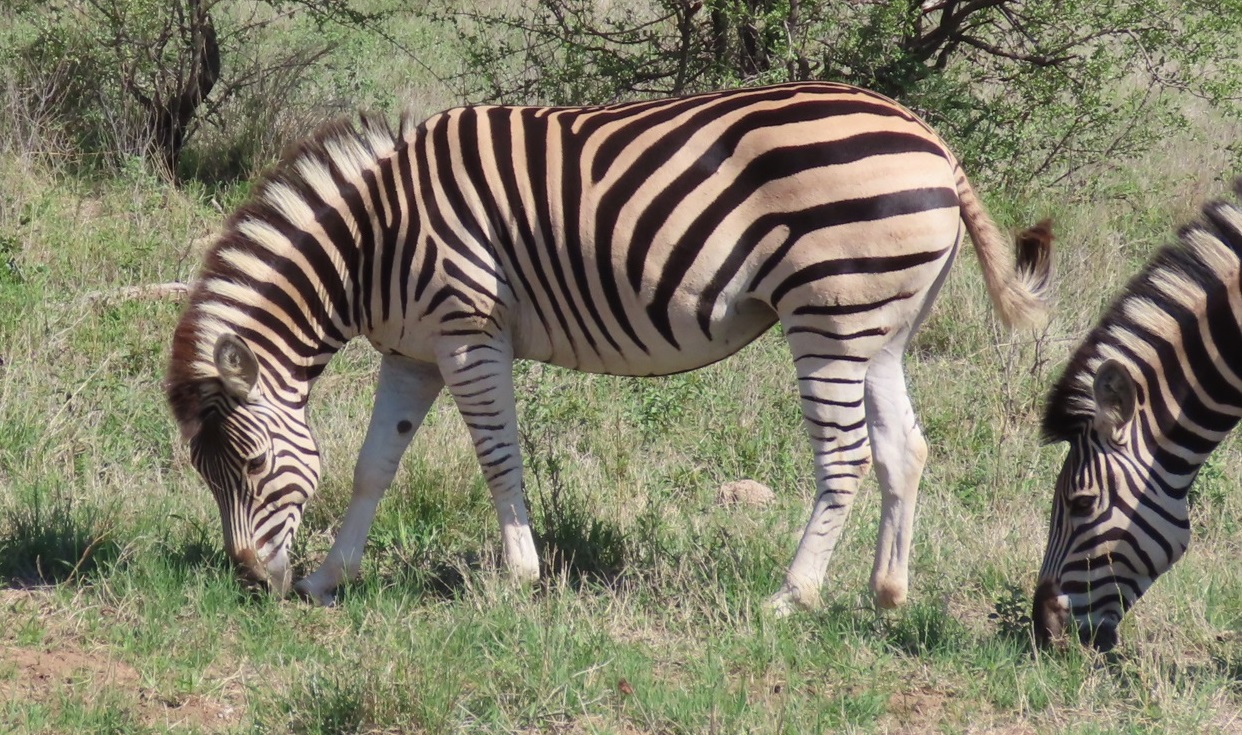}
	\caption[A coordinate atlas]{A non-linear world: animals, plants, trees}
	\label{Fig.1}
\end{figure}
How are these features compatible with the claimed linearly evolving underlying dynamics of a single quantum wave function? 
There is a problem somewhere, even if the wave function lives in a very high dimensional space, or perhaps an infinite dimensional space: there is clearly a basic incompatibility. 

I claim that the solution is that in real world applications, quantum physics applies locally everywhere, but only local wave functions exist; there is no wave function for a cat or similar macroscopic objects such as a brain, let alone for the entire universe.

\subsection{Linearity and non-linearity of quantum theory}
\textbf{Essential linearity of Quantum theory
}

The time-dependent first order Schr\"{o}dinger %/ Dirac 
equation %\cite{Laughlin and Pines}
\begin{equation}\label{eq:Schr}
i \hbar \frac{\partial|\Psi(t)\rangle}{\partial t} = \hat{H}\,|\Psi(t)\rangle
\end{equation} 
for a %many-particle
 wave function $|\Psi(t)\rangle$ (\cite{Laughlin and Pines}, see also
  \cite{Isham}, eq.(1.6); \cite{QM}, eq.(2.8b)),   
is linear in $|\Psi(t)\rangle$ because it is source-free, and the Hermitian  operator $\hat{H}$, which acts linearly on $|\Psi(t)\rangle$, does not itself  involve $|\Psi\rangle$. Thus for arbitrary constants $\alpha$, $\beta$, we have the linearity relation 
\begin{equation}\label{eq:linear}
 i \hbar \frac{\partial}{\partial t}\{\alpha\,|\Psi_1(t)\rangle+\beta\,|\Psi_2(t)\rangle\}  = \alpha\hat{H}\,|\Psi_1(t)\rangle+ \beta\hat{H}\,|\Psi_2(t)\rangle\,.
\end{equation}  

 Key features of quantum theory result  from this linearity \cite{Ellis (2012) QM}:
\begin{itemize}
	\item The use of a complex Hilbert space formalism \cite{Isham}, 
	\item Time reversible Hamiltonian evolution if $\hat{H}$ is time independent  \cite{Cappellaro}, 
\item Interference  between quantum entities, as in the 2-slit experiment \cite{Feynman lectures},
\item Superposition of quantum states  
\cite{superposition}, \cite{Silverman},
\item The possibility of coherence \cite{coherence},
\item Entanglement (\cite{Isham} \S 8.4), \cite{entangle},   
\item Unitarity of the S-matrix in scattering processes \cite{Weinberg},
\item The possibility of using a path integral approach \cite{Feynman and Hibbs},
\item Fermi-Dirac statistics (\cite{QM}:134-137), resulting from the Pauli exclusion principle \cite{Dirac},
\item Bose-Einstein statistics \cite{Ziff}.
\end{itemize}

\textbf{Essential and inessential nonlinearities of quantum theory}
There are of course also some non-linear features of quantum theory. I classify these as inessential or essential.
 
\textbf{Inessential nonlinearity}: Non-linearities in the Hamiltonian $\hat{H}$ do not disturb the linearity relation (\ref{eq:linear}). On expanding $e^{iH}$ in a power series, these non-linearities result in the possibility of using  Feynman diagrams \cite{Feynman and Hibbs} as a calculational tool to solve (\ref{eq:Schr}). As the name implies, the `virtual particles' represented in such diagrams are not real particles, they are rather a conceptual aid in carrying out calculations. Despite the complexity of these diagrams,  
the wave function is still propagated linearly. 

\textbf{Essential nonlinearity 1:}  Unless the initial state is already an eigenstate, wave function \textit{collapse to an eigenstate} takes place  (\cite{Isham}: \S8.3, \S8.5), with probabilities of a specific eigenstate and associated eigenvalue being given by the Born Rule (\cite{QM}:  pp.43-44, \cite{Atkins QM}: p.23). This process is non-linear and irreversible,  so it cannot be described by the linear equation (\ref{eq:Schr}) and information is generically lost \cite{Ellis (2012) QM}. This happens all the time in the real world as photons impinge on  a screen or on a  CCD in a camera or on chlorophyll or rhodopsin molecules, as radioactive atoms decay, as chemical reactions take place, etc. It occurs in laboratory situations when measurements are made, but it is not generically  associated with such a context: it need not be a `measurement' in that sense.
 
Some presentations of quantum theory simply ignore this non-linearity, claiming that quantum theory is always unitary; but without it there is no point in knowing the wave function, as then no experimentally determinable outcomes are predicted by the theory.

There are alternatives that try to avoid this non-linearity, particularly a \textit{many worlds} view (the Everett interpretation) where such collapse never happens (\cite{Isham}: \S8.5.4), sometimes extending to a `many minds' view, often proclaimed in popular expositions. 
Why is this proposed? - it's an extended attempt to claim that only linear dynamics (\ref{eq:Schr}) occurs \cite{Carroll (2022)}. However then deriving the Born rule for probabilities of outcomes, which we determine experimentally to be the case in the real universe, is problematic. A key  technical problem for these theories is the non-uniqueness of relative states (\cite{Isham}:157-159). 
  It is also in my view inconceivable that registering a single photon on a single screen in one laboratory  splits the entire universe in two, right up to the  Hubble radius (and beyond?). Any `many minds' interpretation (\cite{Isham}:157) \cite{Butterfield} is unviable because, apart from any other problems, there is no single wave function for a brain (see below). Tegmark's many varieties of multiverses are critiqued in \cite{Butterfield_1}. 
  
There are also \textit{hidden variable/pilot wave} theories (\cite{Isham}:160), \cite{Valentini} that avoid wave function collapse, with randomness of outcomes being due to the unknown values of hidden variables. They were initially crafted so that their outcomes would be identical to those of standard quantum theory (\cite{Bohm}:110), but Bell's work led to a variety of proposed tests of those theories %in situations with outcome different from the standard theory,
 e.g. \cite{Zeilinger},  \cite{Wiseman Bell}. However if these variables do indeed exist, we have no access to their values %before outcomes take place
  except immediately after an experiment (when also the standard theory gives unique outcomes: the same eigenvalue and eigenvector as just  measured). Their postulated existence makes no difference to the quantum mechanics impossibility of generically making unique predictions  of experimental results from determinable initial data %because generically 
  (``\textit{One cannot even in principle know the hidden variable values}''  \cite{Genovese}:376). %  (generically we can retrodict to determine their values to earlier times, but that does not help). %  (when standard theory also gives a unique result: the same eigenstate as just determined). %But this is the basic aim of physics.
  %It is also unclear as to how those hidden variables could have initial values such as to lead to biological emergence %, for example
  %- existence of zebras, or %outcomes such as  
  %specific rational thoughts like  Einstein's discovery of General Relativity Theory. %: the cosmological context  does not have such data in it %, modulated random Gaussian fluctuations occur
  % on the Last Scattering Surface.
  In any case, these theories  do not feature in standard quantum chemistry texts \cite{Buyana} \cite{Kohn 1999} \cite{Atkins QM}, \cite{Karplus muotiscale}, which develop from the standard approach \cite{QM}, so I will not consider them further.
%How did those hidden variables get to have values leading to these outcomes?

I instead support the idea of classical outcomes arising by \textit{Contextual Wavefunction Collapse} \cite{Drossel and Ellis quantum}. This is in essence  a version of the Copenhagen interpretation where the measurement apparatus can be regarded as classical, as experienced in practice by physicists \cite{Bub}. It is supported by Bohr's idea of the property of a quantum system having a meaning only within the context of a specific measurement situation, as suggested also by the Kochen-Specker Theorem (\cite{Isham}:p.167), as elaborated recently in \cite{Landsman}.  
 Real quantum measurements are complex multi-stage affairs that are crucially context-dependent  (\cite{Drossel and Ellis quantum}:\S2), \cite{Drossel 2023}. 

\textbf{Essential nonlinearity 2:} The dual process to measurement is \textit{state vector preparation}, which is also non-unitary.  The key  feature of quantum state preparation is pointed out by Isham as follows (\cite{Isham}:74,134): selected states are drawn from some collection $E_i$ of initial states by a suitable apparatus, for example selected to have some specific spin state, as in the Stern-Gerlach experiment; the other states are discarded. Again it is a non-linear contextual process due to the nature of the relevant apparatus (\cite{Ellis (2012) QM}:\S.5.2.6). 

\textbf{Essential nonlinearity 3}: The square in Born's Rule is a basic quantum non-linearity: the  wave function does not directly give probabilities, but by a square (\cite{Isham}:5, \cite{QM}:43-44). This is a key difference between classical and quantum physics, and is essentially why complex numbers are needed in quantum theory \cite{complex}.
\\

\textbf{Real world quantum linearities}
Famously, both superposition and entanglement have been demonstrated at macro scales. However to do so, one must prevent non-linear effects happening,
and so need a very carefully engineered context:
surfaces must be precisely machined, interactions with heat baths must be minimized, very low temperatures are usually needed at considerable cost, decoherence must be fought by isolation.

 This is quite different than conditions in a biological system such as a living cell \cite{The Cell}, or a brain made  of networks of billions of cells joined by synapses \cite{Kandell}. No linearity (quantum or otherwise) applies in these contexts.

\subsection{Real world macroscopic non-linearities}
As emphasized at the start, the real world is highly non-linear at emergent scales. 
A key example 
is  \textbf{\textit{feedback control loops}} (Figure 2).
\begin{figure}[h]
	\centering
	\includegraphics[width=0.6\linewidth]{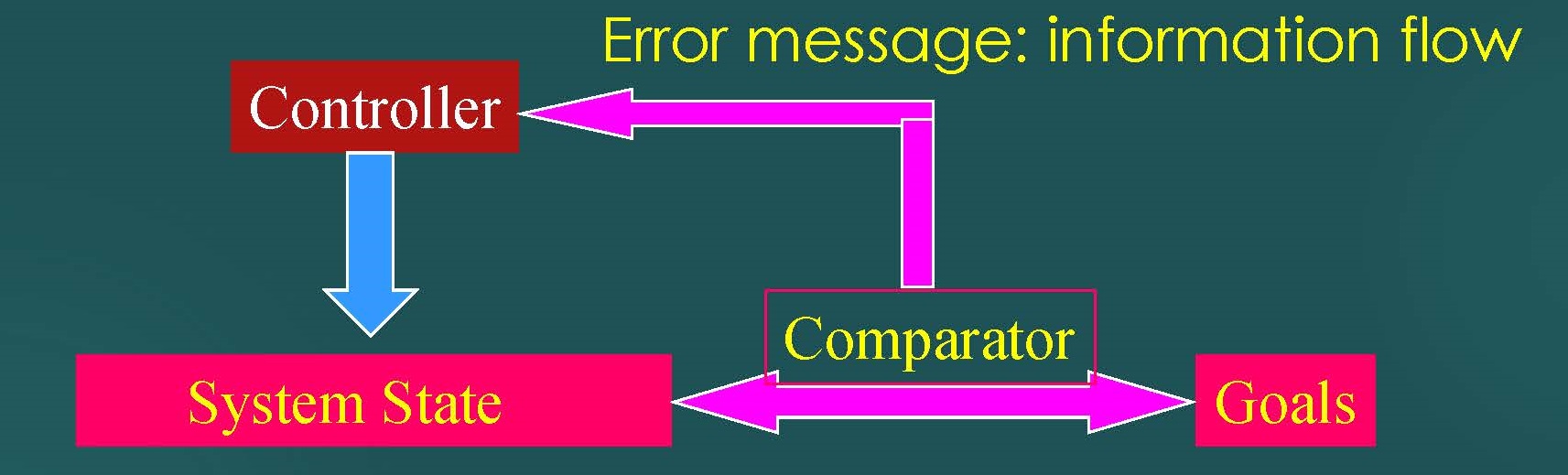}
	\caption[Feedback control loop]{Feedback control loop}
	\label{Fig 2}
\end{figure} Thousands occur at all levels in biology, where they underlie homeostasis \cite{Homeostasis}, for example controlling body temperature, blood pressure, electrolyte levels, and so on. These systems are key to biological function: you are ill if the relevant variable is out of range. 
They are central to much engineering (e.g. aircraft autopilots and chemical engineering process control) where they are labeled  \textit{cybernetic systems} \cite{Wiener} \cite{Ashby}. Daily life examples are thermostats for temperature control of water in a water cylinder, or of air in a room.

The general nature of feedback control is shown in Figure 2. A system goal $G$ is set, and a comparator compares the actual system state $S$ with that goal. If they differ, a classical error message $E$ is sent to a controller that takes action to correct that error. Information flows through specifically constructed feedback loops, enabled by the emergent structure of the 
system. 
In the case of a thermostat for a room, the  goal is a chosen comfortable temperature $T_0$ set on the thermostat controller. The control circuit in Figure 2 is realised by electrical wiring that either sends current to the heater, or not, depending on the sensor state; this is a carefully engineered emergent structure based in the underlying atoms. It's nature cannot be captured by coarse graining, as is the case with the kinetic theory of gases.  
                      
The outcome is not linear, rather branching dynamics of the following form occurs: 
\begin{equation}\label{eq:Temp}
\{\textrm{IF}\,\, (T < T_o) \,\textrm{THEN}\, (apply\, heat)\, \textrm{ELSE}\, (not)\} \Rightarrow \{T \rightarrow T_0\,> 0\}.     
\end{equation}
which cannot occur for a linear source-free 1st order ODE \cite{odes}. Key points are,
\begin{itemize}
	\item The initial data is irrelevant to the  outcome: it is determined by the goal $T_0$;
	
	\item The branching dynamics (\ref{eq:Temp}) is determined physically by time-dependent constraints \cite{Juarrero}, namely the way the wires in the electric circuit control the flow of electrons, depending on the temperature $T$.  
	\item If heating is needed, the topology of the circuit is closed and current flows. If it is not needed, the circuit is open and no current flows.The emergent structure thus results in discontinuous dynamics. 
	\item Hence it is not unitary dynamics :  initial states $T_1<T_0$ and $T_1' = (1/2)T_1$ lead to the same outcome $T_0 > 0$ in a finite time. %contrary to (\ref{eq:linear}), which implies the outcomes should differ by a factor $1/2$.
	\item It therefore cannot be described by linear evolution (\ref{eq:Schr}) of a single wave function $|\Psi(t)\rangle$ 
	\item If the topology of the emergent structure is changed by either disconnecting the wires to the sensor, or reversing them, the outcome is different. In the first case, no homeostatic action takes place; in the second case, the system burns out. All the components are unchanged.
	\item The setting of the thermostat determines the average speed of motion $\langle v^2\rangle$ of molecules in the room, because they are determined by the temperature $T$. Thus this is top-down causation from the emergent macrostructure of the thermostat to the molecules in the room.
\end{itemize}
A second key example of emergent non-linear systems in biology is contextually-sensitive  \textbf{\textit{oscillators}}, for example in the heart \cite{Noble heart (2012)}, with the function of pumping blood and thereby keeping cells alive; thalamocortical oscillations \cite{Thalamocortical oscillations}; and adaptive resonant circuits in the brain, underlying attention, learning, and recognition \cite{Grossberg}.

\begin{figure}[h]
	\centering
	\includegraphics[width=0.5\linewidth]{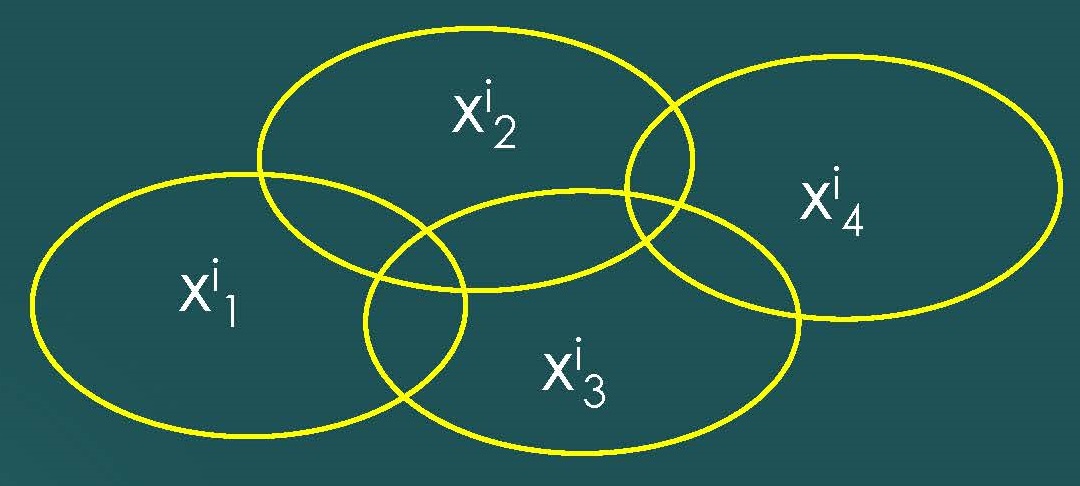}
	\caption[A coordinate atlas]{A coordinate atlas for local coordinates $x^i_N$.}
	\label{fig:3}
\end{figure}

\subsection{Atlases and local wavefunctions}
How to reconcile local unitary dynamics (\ref{eq:linear}) with emergent branching dynamics (\ref{eq:Temp})? The key idea is \textit{local quantum wavefunctions.} To set the scene I will first introduce the idea of local coordinates as used in in General Relativity theory. 

\textbf{Prolog: GR key idea: Local coordinates}. 
The revolution of global studies in General Relativity Theory \cite{HE} arose via the concept of \textit{local coordinates} covering part of a manifold, joined via overlap domains (Figure 3). 
An atlas covers the manifold globally, but often no single coordinate system exists for the  whole: e.g. this is not possible even in the case of the 2-sphere $S^2$. Use of atlases made global GR studies possible, in particular the study of horizons in black holes via  various coordinate systems and the transformations between them. 

\textbf{QM Resolution: Local wave functions.}
Proposal: Essentially the same applies to quantum theory:
 Local wave functions exist everywhere, but no global wave function exists, nor even a single wave function for most macro objects, e.g. a living cell, a cat, or a brain.

\begin{figure}[h]
	\centering
	\includegraphics[width=0.55\linewidth]{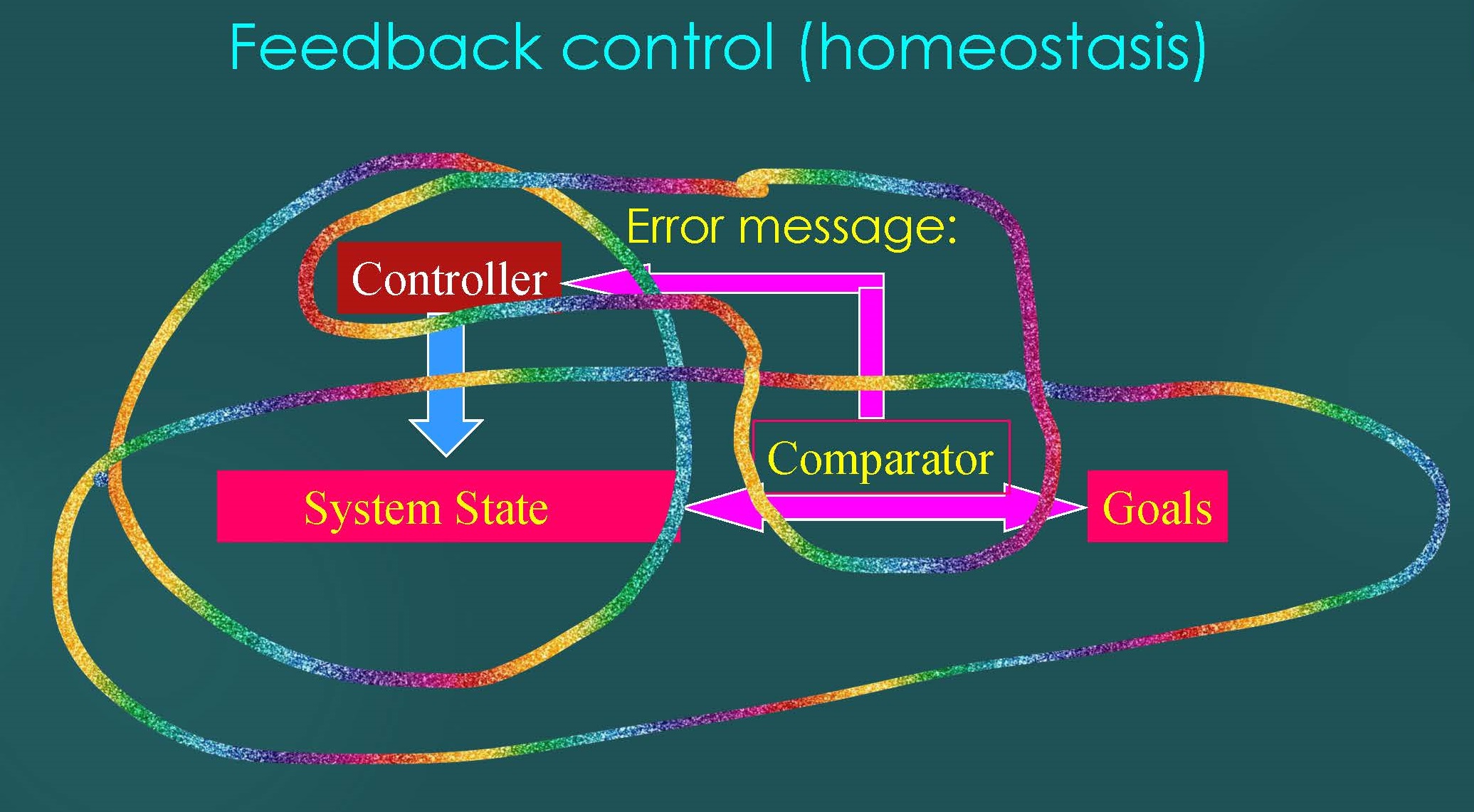}
	\caption[A coordinate atlas]{Local wave functions for a feedback circuit}
	\label{Fig.4}
\end{figure}
\textbf{Example: Feedback control (homeostasis)} as in Figure 2. In this case, local wave functions  would exist and underlie the dynamics in each of the three domains shown in Figure 4, where the dynamics is linear. The fundamental non-linearity of the feedback loop arises by combining the linear dynamics of local wave functions in each of these domains. Thus there is generically a patchwork of local wavefunctions valid in local domains, as in Figure 4.
Combined, these cover the system globally, but no single wave function exists for the  whole

Each cell in your body has thousands of feedback loops \cite{The Cell}. The dynamics is highly non linear. There is no way a cell can be described by a single wave function with linear dynamics (\ref{eq:Schr}). But local wavefunctions as in Figure 4 will do the job.

\textbf{Example: Diffraction of extended bodies} A physics example of this approach is the case of diffraction of extended bodies, presented in \cite{Schutz extended}.

\textbf{Heat Baths} are fundamental to describing macroscopic systems, but cannot be described by a many particle wave function \cite{Drossel 10 reasons}. From the viewpoint just put, the local wave function domains become so small and so dynamic that it no longer makes sense to separate them; one should rather use the appropriate classical concepts. \cite{Drossel condensed} explains:
 \begin{quote}
 	\textit{``The Schr\"{o}dinger equation for a macroscopic number of particles is linear in the wave function, deterministic, and invariant under time reversal. In contrast, the concepts used and calculations done in statistical physics and condensed matter physics involve stochasticity, nonlinearities, irreversibility, top-down effects, and elements from classical physics. This paper analyzes several methods used in condensed matter physics and statistical physics and explains how they are in fundamental ways incompatible with the above properties of the Schr\"{o}dinger equation. The problems posed by reconciling these approaches to unitary quantum mechanics are of a similar type as the quantum measurement problem. This paper, therefore, argues that rather than aiming at reconciling these contrasts one should use them to identify the limits of quantum mechanics. The thermal wavelength and thermal time indicate where these limits are for (quasi-)particles that constitute the thermal degrees of freedom.''}
 \end{quote}
This view is strengthened by the fact that a cell is a very dynamic context: topological turbulence occurs in the membrane of a living cell \cite{top turb}, molecular machines extract order from chaos \cite{Ratchet}, and the brain is an essentially noisy organ \cite{Rolls and Deco}.\\

\textbf{Summary}

1. Quantum physics applies locally everywhere via local wavefunctions, but does not apply globally with a single wavefunction.

2. Local wavefunctions are restricted to domains where the dynamics is linear \cite{Ellis (2012) QM}. 

3. This split cannot be characterised simply in terms of energy or length scales: it is contextually dependent, for example sufficient isolation of a system to avoid decoherence. 

4. Non-linear dynamics arises by combining local  linear dynamics in non-linear ways.

\subsection{Local wave functions imply macro objects are classical}
This view supplies a suitable context for interpreting what Leggett states:
\begin{quote}
	``\textit{It is quite conceivable that at the level of complex, macroscopic objects, the quantum mechanics superposition simply fails to give a correct account of the dynamics of the system}'' (\cite{Leggett}):98).
\end{quote}
Isham confirms this possibility (\cite{Isham}:149): 
\begin{quote}
	``\textit{A macroscopic subsystem (or ensemble of such) may not be describable by any state at all, not even a linear superposition of eigenstates}''.
\end{quote} 
\begin{figure}[h]
	\centering
	\includegraphics[width=0.6\linewidth]{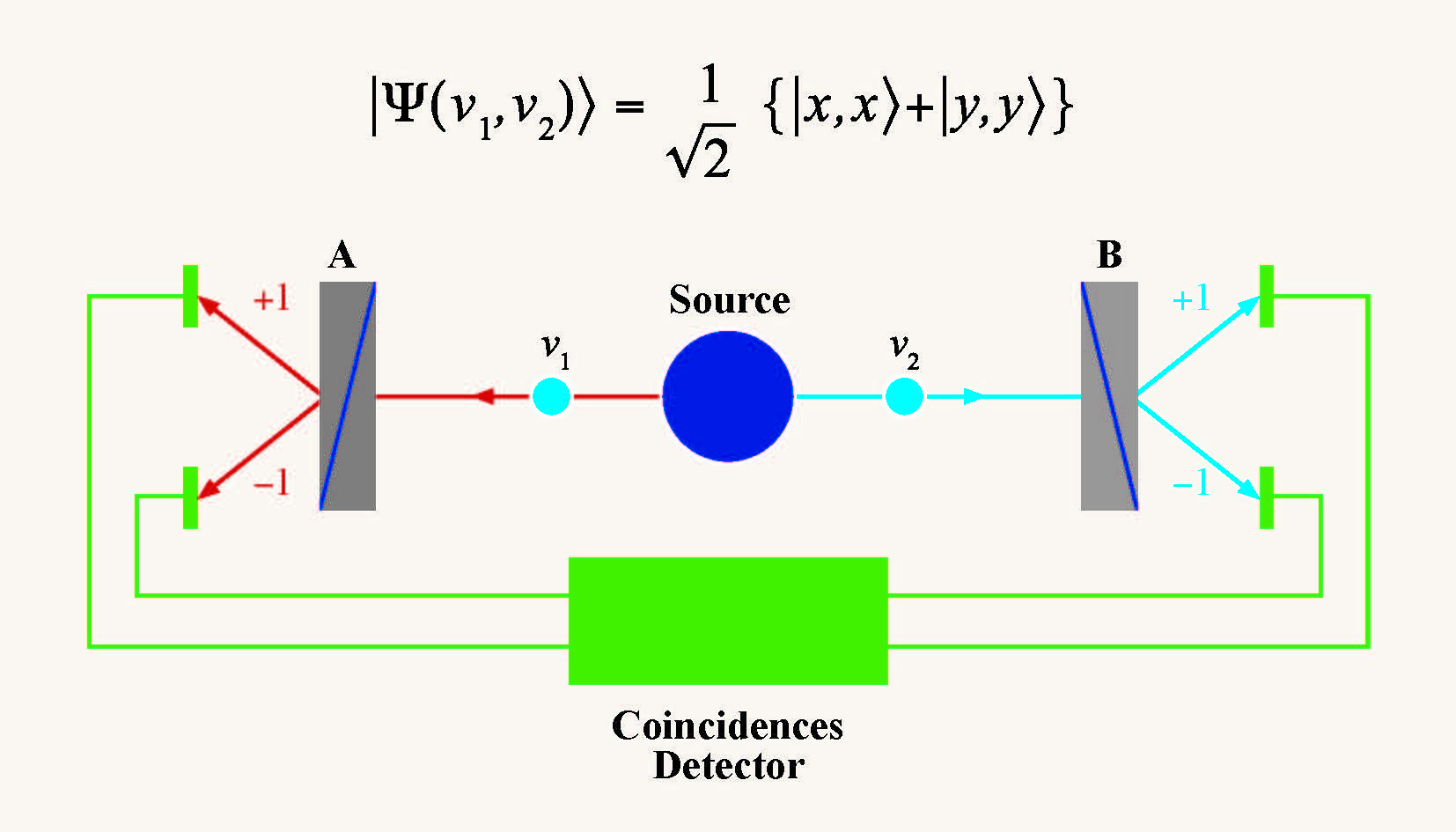}
	\caption[Bell test by Alain Aspect]{Bell test: Alain Aspect (adapted from \cite{Aspect (2015)} by Mandy Darling).}
	\label{Fig.5}
\end{figure}

This leads to a defensible version of the Copenhagen Interpretation of quantum theory. Any measuring apparatus is classical, as in the traditional view supported by Bohr: ``\textit{The measuring device is located firmly in an external classical world, thereby opening an irreducible gap between the quantum system and the instrument of observation} (\cite{Isham}: p.132). One cannot describe it in terms of a single wave function because its dynamics are not linear, and  because as it is macroscopic,  in thermodynamic terms it contains effective heat baths which cannot be represented by a multiparticle wavefunction \cite{Drossel 10 reasons}, \cite{Drossel condensed}.

\textbf{Example:  Aspect's paper on the Bell test} \cite{Aspect (2015)}, see Figure 5.
All is classical: the source, beam splitters, detectors, 
coincidence detectors),  except a few
interacting particles ($\nu_1$ and $\nu_2$). All the rest are conceived of and described in classical terms. To describe them in quantum terms would make the analysis impossible (\textit{inter alia}, the elements of the apparatus would have no definite state).

\textbf{Schr\"{o}dinger's cat} In particular \cite{Silverman rabbits}, there is no single wave function for a cat as a whole, made of many living cells . As stated in \cite{Silverman} (p.26)):
\begin{quote}
	``[..] \textit{not everything can be ascribed a wavefunction. Cats, rabbits, chickens, and other ludicrous examples of flora and fauna, which have been employed in popular accounts of quantum mysteries, do not have wave functions} .'' 
\end{quote}
Thus because there is no wave function $|\Psi_{cat}\rangle$,  the proverbial 
\begin{equation}\label{eq:alive_dead}
|\Psi_{cat}\rangle \,= \,\alpha|\Psi_{alive}\rangle +\beta|\Psi_{dead}\rangle
\end{equation}
is not a legitimate equation, as none of these wave functions exist! The only quantum feature in a typical diagram of the Schr\"{o}dinger's cat setup is the set of excited atoms that trigger the detector in a random way.  Everything else - the cage, the detector, the poison vial, the hammer that breaks it - is classical, including the cat.

\subsection{Is there a wave function for the universe?}
No, there cannot exist single wave function $|\Psi_{universe}\rangle$ for the universe as a whole, because heat-baths and cats exist in the universe, as do vast numbers of living cells, each containing thousands of feedback loops. 

This statement undermines a vast amount of writing concerning alleged outcomes of the existence of such a wavefunction. Of course the idea was originally applied  to minisuperspace, where only cosmological degrees of freedom were considered  \cite{Minisuperspace}.  This is a far cry from the way it is used today, with $|\Psi_{universe}\rangle$  supposed to apply to everything including brains, and there is much talk of minds splitting in two because of the Everett interpretation. 

\textit{Astrophysical and idealised black holes} Is there a single wave function for a black hole? 
 In the case of realistic astrophysical black holes, there are many non-linear processes taking place, with heat baths involved \cite{Begelman and Rees}. The answer will surely be no. Whether there can be a single wavefunction for an idealised black hole, as envisaged in the huge literature on the ``information paradox'',  would seem to be a moot question. From the viewpoint of this paper, that assumption needs to be justified.

\section{The puzzle of complexity} 
How does this all relate to complex systems such as life, digital computers, aircraft, cities? 
These are all associated with function or purpose,  whereas there is no purpose in physics \cite{Hartwell et al}, (\cite{NAS Physics of Life}:17,47-48). 
How does this arise out of physics? Via conformational properties of macromolecules in modular hierarchical structures: change of shape
of macromolecules, enabled by quantum chemistry effects \cite{Karplus muotiscale}, underlies biological function at higher levels \cite{Lehn (2004)} \cite{Lehn 2007}. Time dependent constraints enable branching dynamics to emerge, as in the case of feedback control loops (eqn.(\ref{eq:Temp})).

\subsection{The Hierarchy of Structure and Causation}
Complex systems such as life, digital computers, aircraft, cities, are  \textit{multiscale adaptive modular hierarchical structures} \cite{Ellis Topdown} \cite{multiscale}, where each word is important.

\textbf{Why structures?}
Because that is how desired function emerges \cite{function}: eyes enable seeing, lungs enable breathing, and so on. Even proteins have functions enabled by their structure, see \textit{Protein Structure and Function} \cite{Petsko and Ringe}. One can alternatively emphasize the processes carried out by these structures, which are indeed central to function  \cite{Dupre}  - but no processes can occur  without the structures.

\textbf{Why modular?}
The basic principle for handling a complex task is, break up the complex task into simpler tasks; create modules - semi-autonomous parts - to handle the simpler tasks; then merge the outcomes to complete the complex task. This allows abstraction, information hiding, and encapsulation \cite{Booch (1990)}. It furthermore allows evolutionary development of modules, with inheritance: existing modules are adapted, with modifications, to fulfill new functions. Indeed this is the only way to develop truly complex systems \cite{Simon}.

\begin{figure}[h]
	\centering
	\includegraphics[width=0.55\linewidth]{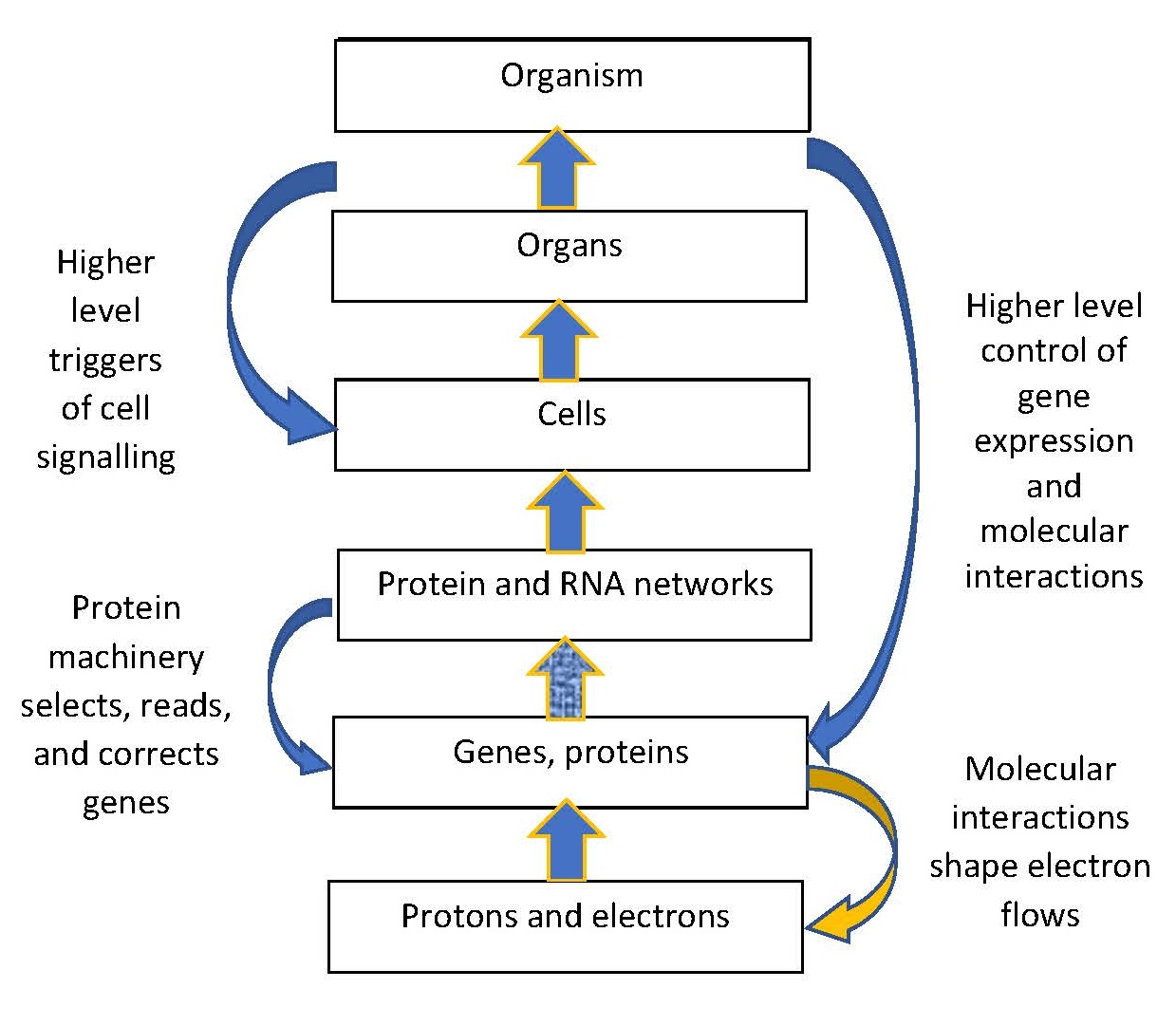}
	\caption[Branching Physics in a Biological Context]{\textbf{Branching Physics in a Biological Context} \textit{Branching biological logic chains down to cause the underlying physical dynamics to branch} From \cite{Ellis and Kopel}.}
	\label{Fig.6} 
\end{figure}
\textbf{Why hierarchical}? Emergent levels, each with a different nature,  naturally result on combining modules to give a multiscale structure.   The hierarchy of emergence for biology \cite{Rhoades and Pflanzer} \cite{multiscale} is shown in Figure 6. There is a higher level of society, and lower levels of underlying physics (quarks, gluons; maybe strings). For practical purposes, the level of protons, neutrons, and electrons is the lowest level that counts for everyday life \cite{Laughlin and Pines}.
Each level handles different kinds of tasks, supporting other levels. Each level is needed for the whole to work, which is the true nature of causal closure \cite{Montevil and Mossio closure} \cite{Ellis Causal closure}. Each level is equally significant for the functioning of the whole; none is more fundamental than any other \cite{Noble (2012)}.

The hierarchy is \textbf{multiscale} over many orders of magnitude,  with huge numbers of particles constituting huge numbers of cells that together make up the whole  \cite{multiscale}.

 \textbf{Multiple realisability} takes place as causation chains down from higher to lower levels: many lower level elements can give essentially the same higher level outcome, and this degeneracy is huge at the genotype level \cite{Wagner arrival}. Relation to function is hidden at this level.

\textbf{Why adaptive?} The system must be able to respond effectively to the environment, which will generically be of a dynamic nature and require different responses at different times and places in order to navigate it safely, obtaining energy and information as required \cite{NAS Physics of Life}.

 \subsection{Emergence of higher effective levels}
Emergence takes place over three different timescales: short term (function), medium term (development), and long term (evolution). Multiple realisability applies in all these cases.

\textbf{Functionally},  different descriptions, variables, and effective laws apply at each emergent level at each time, because symmetry breaking occurs between levels \cite{Anderson 1972}. In functional terms, 
  the levels are linked by \textit{upward emergence} due to coarse graining, explicit symmetry breaking,
 or black boxing, and by \textit{downward causation} \cite{Ellis Topdown} due to time dependent constraints \cite{Juarrero}, together with the ability of higher levels to create, modify, or delete lower level elements (`machresis', \cite{machresis}), leading to the idea of ``understanding the parts in terms of the whole'' [Cornish-Bowden \textit{et al} (2014)], as for example in proteostatic regulation in neuronal compartments \cite{Giandomenico}. Here \textit{causation} is defined as difference making \cite{difference}. 
 
  These dynamics together enable  \textit{same level causation} to occur at every emergent level \cite{Noble (2012)}, expressed by effective laws that apply at that level \cite{Ellis Emerge SS and biology}.
  These laws generically exhibit contextually determined logical branching  of the form, 
 \begin{equation}\label{eq:bio branching}
 \textrm{IF} \,\, T(X)\,\, \textrm{THEN}\,\,  F1(Y)\,\,  \textrm{ELSE} \,\, F2(Y),
 \end{equation}
 where $Y$ is a dynamic variable at level $L$
 and $X$ is a control variable at level $L$ or level $L+1$. Such contextual branching is the key difference between physics and biology \cite{Ellis and Kopel}, enabling function to emerge \cite{Hartwell et al}. 
 Homeostasis such as (\ref{eq:Temp}) is a case of such branching due to  time dependent constraints. Metabolism \cite{metabolism} is an example of machresis in physiology and cells, as are brain plasticity effects associated with memory via gene regulation \cite{Kandel} which determine what proteins will be present.
 
\textbf{Developmentally} emergence takes place over timescales of between minutes to decades as an organism develops from a single cell to a cooperative ensemble of $10^{11}$ to $10^{13}$ cells of different types. The nature of the cells present at each place in the organism is determined by developmental processes \cite{Wolpert} controlled  by positional information conveyed by morphogens in the ecological context \cite{Gilbert development}. This is a crucial form of machresis.

\textbf{In evolutionary terms}, 
natural selection \cite{Darwin} \cite{Mayr Evln} takes place over geological or shorter timescales, depending on the reproduction timescale.  Reproduction with variation, followed by natural selection of phenotypes with a relative reproductive advantage in a specific ecological context,  leads to selection of genes producing proteins that result in existence of phenotypes better suited to the environment, naturalistically explaining apparent design. The selection principle is relative reproduction success. 

Downward causation takes place because different environments result in different outcomes \cite{Campbell 1974}. Thus for example bears have adapted differently to Arctic snow (polar bears) vs Canadian forests (brown and black  bears), so that they will in each case be able to hunt without being conspicuous. Thus these different contexts result in different genes (DNA)
resulting in changed proteins resulting in different fur colour. The selection causal chain is 
\begin{equation}\label{eq:select}
Environment \Rightarrow Selected\, Organisms  \Rightarrow Selected\, Proteins\,\Rightarrow DNA\, sequence  .
\end{equation}
Other examples are given in \cite{Wagner arrival}: ecological advantage acts down to select preferred proteins via associated genes.  This is clearly multilevel selection, with a crucial feature:
 \begin{quote}
 	 \textit{Selection shapes all emergent levels $L_I$ in Figure 6 simultaneously. It has to do so, because they all work together to enable the organism to function \cite{Noble (2012)} \cite{Ellis Causal closure}. Selection is not confined to either the gene level or the organism level.}
 \end{quote}
 Three closing comments follow, regarding neutral evolution, the `Evo-Devo' viewpoint, and major evolutionary transitions. 
 
\textit{Neutral evolution} Essentially as a result of the huge multiple realisability occurring in the  process (6), some population geneticists claim that non-adaptive neutral evolution, resulting from genetic drift, almost always takes place rather than adaptive selection \cite{Kimura}, \cite{neutral}. 
Underlying this view is a definition of ``function'' based in selection of DNA segments (``\textit{Functional DNA is DNA that is currently under purifying selection - it is being maintained by natural selection}'' \cite{Moran}), rather than defining function in terms of physiological or ecological relations, as in   \cite{Hartwell et al}, \cite{Ellis Emerge SS and biology}, \cite{NAS Physics of Life}.  

 This results in the paradox of these workers in effect denying the central achievement of evolutionary theory, namely the explanation of apparent design \cite{Gardner adaptation}. Evolutionary theorists studying evolution in relation to physiology, ecology, or behaviour do not take this view (see e.g. \cite{Wagner arrival}).  This paper adopts the organisational  definition of function \cite{Mossio et al}, \cite{Farnsworth},   (\cite{Ellis and Kopel}:\S1.3).

\textit{Evo-Devo} In reality, evolutionary and developmental aspects interact with each other. It is developmental systems that make it possible for an organism to exist \cite{Gilbert development}, so it is they in particular that are selected by evolutionary processes \cite{Griffiths developmen systems}. Their action then alters evolutionary outcomes, leading to the Evolution-Development interaction (Evo-Devo) view of the dynamics taking place \cite{Carroll Endless Forms}, \cite{Carroll EES}. In this way gene regulatory networks are selected that enable more complex logic (AND, OR, NOT, and so on) to emerge \cite{Jacob and Monod}, \cite{Monod et al} via cell signalling processes \cite{Berridge signalling}.

\textit{Major transitions in evolution} \cite{major transitions} correspond to new ways of transmitting information between generations, perhaps new levels of cooperation (entities that were capable of independent replication before can only replicate as part of a larger whole after), and in some key cases, adding a new level $L_J$ to the hierarchy.

\subsection{The key physics-biology link: Supramolecular chemistry}
The contextual control of supramolecular shape \cite{Lehn (2004)}  is the key physics-biology link.
\begin{figure}[h]
	\centering
	\begin{minipage}{0.45\textwidth}
		\centering
		\includegraphics[width=0.55\linewidth]{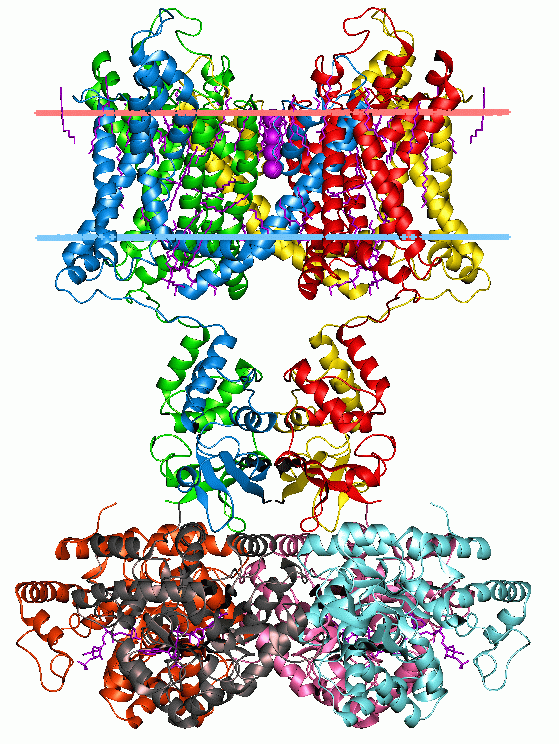}
		\caption{Membrane potassium ion channel structure %in a membrane-like environment,
			when the channel is closed. \textit{The  channel shape alters according to the voltage difference across the membrane, hence allowing or (in this case) impeding ion passage.}
			%Calculated hydrocarbon boundaries of the lipid bilayer are indicated by red (top) and blue (lower) dots.} 
			Diagram by Andrei Lomize. From the Open Membranes %(OPM)
			Database. %https://opm.phar.umich.edu/contact
			%with permission  %\href{https://en.wikipedia.org/wiki/Voltage-gated_potassium_channel}{Wikipedia}.
		}  
		\label{fig:7}
		\quad 
	\end{minipage}\hfill
	\begin{minipage}{0.45\textwidth}
		\centering
		\includegraphics[width=0.55\linewidth]{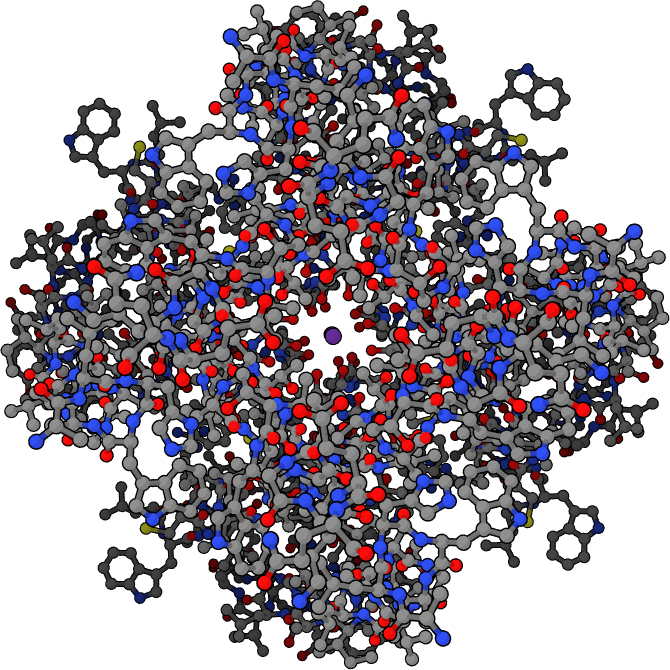}
		\caption{Membrane potassium ion channel structure when the channel is open \textit{Top view of potassium ion (purple, at centre) moving through the potassium ion channel to the cell interior when the channel is open, because the hinged  paddles have moved outwards.} (Protein Data Bank:1BL8,  open access)}
		%https://www.ebi.ac.uk/pdbe/._channel}{Wikipedia}}
		\label{fig:8}
	\end{minipage}
\end{figure}

\textbf{Voltage gated ion channels} An important example is the behaviour of voltage gated ion channels in axons. This underlies the Hodgkin-Huxley equations \cite{Hodgkin and Huxley}, resulting in action potential spike chain propagation in neurons \cite{Kandell}.

The key is the changing conformational shape of the protein ion channels imbedded in the cell wall, responding to electric voltage across the cell wall. The channel has four paddles that can change position, seen from the side in Figure 7 and from the top in Figure 8, which move in response to the voltage across the cell wall. They are closed when the voltage gradient is outwards (more positive in the exterior), as in Figure 7, so $K^+$ ions cannot enter the cell. They are open when the voltage gradient is inwards (more positive in the interior), as in Figure 8, so $K^+$ ions can enter the cell.  This is downward causation from the scale of the ion channel to the scale of the much smaller ion, controlled by the potential gradient across the cell wall.  

This context-dependent conformation change enables logic of the form (\ref{eq:bio branching}) to emerge from the underlying physics: 
\begin{equation}\label{key}
\textbf{IF}\,\, (outward)\,\, \textbf{THEN}\,\, (channel\, closed)\,\, \textrm{IF}\,\, (inward)\,\, \textbf{THEN}\,\, (channel\, open).
\end{equation}
This inflow of ions then alters the voltage across the ion channel, so circular causation occurs \cite{Noble et al 2019} and underlies the oscillations that create action potential spike chains.

When this occurs in billions of ion channels imbedded in the neural networks of the brain, it enables propagation of action potential spike chains in neurons \cite{Kandell}, thereby enabling existence of adaptive resonant circuits \cite{Grossberg} and hence underlying the emergent computational brain \cite{Churchland and Sejnowski}, able for example to calculate solutions to the Schr\"{o}dinger equation (\ref{eq:Schr}) via conditional branching (\ref{eq:bio branching}). The logical branching occurring in such thoughts and planning emerges in this way through the modular hierarchical structure of the brain \cite{Churchland and Sejnowski}, (\cite{NAS Physics of Life}:144-152).

Such hugely complex molecules with key biological functions came into existence via evolutionary selection  \cite{Wagner arrival}) because of the crucial role they play in enabling intelligence to emerge, and hence enhanced our ancestor's survival prospects in the past. All proteins have functions \cite{Petsko and Ringe} which have led to their existence via adaptive selection.\\

\textbf{Molecular recognition, logic, and information}
Molecular shape underlies the lock and key mechanism of molecular recognition \cite{Lock and key} \cite{Berridge signalling}  that in turn underlies  function in supramolecular chemistry  \cite{Lehn (2004)}  \cite{Lehn 2007}.
Complex chemical systems arise from components interacting via non-covalent intermolecular forces, underlying the chemistry of molecular information  processing at the supramolecular level  through molecular recognition processes. Cell signalling \cite{Berridge signalling} occurs, underlying gene regulatory networks  and metabolic networks \cite{Goelzer}. Thus  life is  based in information use and logic \cite{Godfrey-Smith} \cite{Nurse}  \cite{Farnsworth Infn}. % 

\subsection{Causal closure} Given this context, causal closure is not a property of the physics level alone, as is often claimed. Because of the confluence of upwards and downwards effects, it occurs only in the context of the interacting whole depicted in Figure 6 via causal closure of constraints  \cite{Mossio et al 2013} \cite{Montevil and Mossio closure}  \cite{Ellis Causal closure}. In the case of digital computers, for example, it includes the reasons why an algorithm was written that is controlling flow of electrons through gates \cite{Drossel and Ellis computers}. In the case of the functioning of biomolecules via the quantum physics of supramolecular chemistry \cite{Karplus muotiscale}, it includes the ecological reason for that molecule existing, for example the functioning of antifreeze proteins that enable the arctic cod to survive in waters that regularly chill below $0$ C (\cite{Wagner arrival}:107).
 The laws of physics by themselves, e.g. Maxwell's equations, do not determine any specific outcomes whatever. Rather they determine the possibility space within which such outcomes arise \cite{Adlam}.
  
 In summary: ``\textit{The complement of reduction is emergence. While reduction considers how the whole can be explained in terms of its parts, the concept of emergence considers the qualitatively new properties of the whole, which are not properties of the parts ... The microscopic level of a system is causally open to influences from the macroscopic environment}'' \cite{Drossel 2023}. 

\section{The role of quantum physics}
How does all this relate to quantum physics? In many ways: it underlies the existence and stability of chemical elements; their nature, as characterised by the periodic table of the elements; the existence and nature of chemical bonds that enable polymer chains to exist and underlie supramolecular chemistry, with its logical branching underlying biological signal processing \cite{Karplus muotiscale}.  Time dependent constraints occur \cite{Juarrero}, so the Hamiltonian is no longer time-independent: $\hat{H} = \hat{H}(t)$ \cite{Ellis and Kopel} and the underlying quantum evolution based in (\ref{eq:Schr}) is therefore no longer unitary, and is not time reversible \cite{Drossel condensed}. 

\subsection{Nature and stability of chemical elements}
Stability of matter arises via the Pauli exclusion principle \cite{Leib stability of matter} \cite{Leib stability}.   
The exclusion principle also underlies the existence of electron shells, and hence determines the \textit{Periodic Table of the Elements}  \cite{Pyykko Periodic table} \cite{Scerri}, which is crucial to all biology.

\subsection{Quantum chemistry: bonds} 

\textit{Chemical Bonds}  are firstly due to electrostatic forces between the valence electrons of atoms (\cite{Buyana}:186-188), resulting in two types of atomic bonds creating molecules: namely covalent and ionic. Secondly, there are two types of bonding due to molecular forces: namely van der Waals bonds and hydrogen bonds. These ultimately arise from the Schr\"{o}dinger equation (\ref{eq:Schr}) applied in a molecular context \cite{Laughlin and Pines} \cite{Atkins QM}.

\subsection{Quantum chemistry: simple molecules}

 For polyatomic molecules, one can write a molecular wavefunction that is a combination of atomic wave functions. Its use to study simple molecules is discussed in (\cite{Buyana}:\S6.)  
 
The description at the microscopic level is based on a Hamiltonian $H$ for nuclei of mass $M_i$ and atomic number $Z_i$ at position $\textbf{R}_i$ and electrons of charge $e$ at position $\textbf{r}_j$   \cite{Laughlin and Pines} (\cite{Philips}:16). It has terms for the kinetic energies of the nuclei and of the electrons, and for the potential energies of the Coulombic interactions of the three kinds: nuclei-to-nuclei; nuclei-to-electrons; and electrons-to-electrons:

\begin{eqnarray}\label{eq:hamiltonian}
	H &=& 
	-\sum _i\frac  {\hbar ^2}{2M_i}\nabla _{{\mathbf  {R}}_i}^2
	-\sum _{j}\frac {\hbar ^{2}}{2m_e}\nabla _{\mathbf {r} _{j}}^{2}
	+ \sum _i\sum _{i'>i}\frac  {Z_{i'}Z_{i}e^{2}}{4\pi \epsilon _{0}\left|{\mathbf  {R}}_{i'}-{\mathbf  {R}}_{[]}\right|} \nonumber\\
	&& -\sum _{i}\sum _{j}{\frac  {Z_{i}e^{2}}{4\pi \epsilon _{0}\left|{\mathbf  {R}}_{i}-{\mathbf  {r}}_{j}\right|}}
	+ \sum _{i}\sum _{{i'>i}}{\frac  {e^{2}}{4\pi \epsilon _{0}\left|{\mathbf  {r}}_{i'}-{\mathbf  {r}}_{i}\right|}}\,.\label{eq:Hamiltonian}
\end{eqnarray}
To derive the wave function from this Hamiltonian the actual molecular structure and electron charge distribution in the molecule, plus a series of approximations, is required \cite{Drossel and Ellis computers}.

The \textit{Born-Oppenheimer} (adiabatic) approximation (\cite{Atkins QM}:258-261) \cite{QC McQuarrie} is used, which assumes that the electrons are at all times in equilibrium with the positions of the nuclei.
The wave function is factorized into an electron part $\Psi_{e}(\textbf{r},\textbf{R})$ for given positions of the nuclei, and a nucleus part $\Phi(\textbf{R})$,
\begin{eqnarray}
	\Psi(r,\textbf{R}) =  \Phi(\textbf{R}) \Psi_{e}(\textbf{r},\textbf{R}) \, ,
\end{eqnarray}
leading to the electron equation 
\begin{eqnarray} \label{electronequation}
	(T_e+V_{ee}+V_{ei}) \Psi_{e}(\textbf{r},\textbf{R}) = E_{e}(\textbf{R})\Psi_{e}(\textbf{r,R})
\end{eqnarray}
and the nucleus equation  
\begin{eqnarray}\label{ionequation}
	(T_i+V_{ii}+E_{core}+E_{e}(\textbf{R})) \Phi(\textbf{R}) =  E\Phi(\textbf{R})  \, .
\end{eqnarray}
This is in fact a mixture of classical and quantum physics (\cite{Drossel condensed}:222), because the nuclei are described classically \cite{Born and Oppenheimer} \cite{Hansen Born Oppenheimer}, rather than just being a solution of (\ref{eq:Schr}). As stated by Hendry, ``\textit{The Born-Oppenheimer wavefunction looks more like the solution to
an altogether different equation: the nuclei are treated classically, and we view the
electrons as constrained by the resultant field.}" \cite{Hendry (2006)}, see also \cite{Bishop 2010}. 

A series of further approximations are made to get useful results, including \textit{Molecular orbital theory} (\cite{Atkins QM}:262-286) and the \textit{Hartree-Fock} approximation  (\cite{Atkins QM}:296-316). However this only works for small molecules, as Kohn  stated in his Nobel lecture \cite{Kohn 1999}, awarded for his work on density functional theory: %cf. below, 
\begin{quote}
	\textit{``Traditional multi-particle wave-function methods when applied to systems of many particles encounter what I call an exponential wall when the number of atoms $N$ exceeds a critical value which currently is in the neighborhood of
		$N_0\simeq 10$ (to within a factor of about 2) for a system without symmetries''}.
\end{quote}

 \textit{Density Functional Theory} (DFT).   Because of this  computational complexity of the many-body Schr\"{o}dinger equation,  a solution is beyond reach for larger molecules. DFT  (\cite{Atkins QM}:317-326) sidesteps this problem by making the electron density distribution $n(r)$
 rather than the many-electron wave function  play a
 central role \cite{DF2014}. Underlying
this is the claim \cite{Hohenberg} that the total energy of the system is a unique functional of the electron
 density alone, so it is not necessary to compute the full many-body wave function. The approach  is a mixture of classical and quantum methods (\cite{Drossel condensed}:223).
 However even this method will not work for macro molecules. 

\subsection{Quantum chemistry: macro molecules} The number of atoms in a macromolecule \cite{Lehn 2007} is vastly larger than 10. To handle their complexity requires \textit{multiscale models}. Karplus states in his Nobel lecture \cite{Karplus muotiscale},
\begin{quote}
	\textit{``To develop methods to study complex chemical systems,
	including biomolecules, we have to consider the
	two elements that govern their behavior: 1) The potential
	surface on which the atoms move; and 2) the laws of motion
	that determine the dynamics of the atoms on the potential
	surfaces. ...although the laws governing
	the motions of atoms are quantum mechanical, the key
	realization that made possible the simulation of the dynamics
	of complex systems, including biomolecules, was that a classical mechanical description of the atomic motions is adequate in most cases.''}
\end{quote}
Thus one does not solve (\ref{eq:Schr}) in this case. A classical background provides a basis for the quantum states.  How does this method relate to biological function? Karplus states \cite{Karplus muotiscale},
\begin{quote}
	 \textit{``First,	evolution determines the protein structure, which in many	cases, though not all, is made up of relatively rigid units that are connected by hinges. They allow the units to move with	respect to one another. Second, there is a signal, usually the binding of a ligand, that changes the equilibrium between two structures with the rigid units in different positions. ...This type of conformational change occurs in many
	 	enzymes as an essential part of their mechanism.''}
\end{quote}
This enables cell signalling \cite{Berridge signalling} to occur, underlying metabolic networks (\cite{Wagner arrival}:\S2), \cite{Goelzer}, gene regulatory networks \cite{Jacob and Monod} \cite{Monod et al} (\cite{Wagner arrival}:\S5), and so developmental systems \cite{Griffiths developmen systems}. Overall this is how information is important  at this level in biology \cite{Nurse}.\\

A central issue here is, how does the classical concept of shape, key to occurrence of these processes \cite{Lehn (2004)} \cite{Lehn 2007},  arise out of quantum theory? This is discussed in (\cite{shape Ramsey}, \cite{Bishop and Ellis}:\S5.2). Molecular shape is generated by the environment \cite{Amman chirality} \cite{Amman gestalts}. Biochemistry cannot be reduced to physics \cite{Bishop 2005} \cite{Bishop 2010} because microbiology interactions based in molecular shape is key  in the molecular biology of the gene \cite{Watseon gene} and the cell \cite{Alberts cell}. These emergent processes reach down to coordinate  interactions of proteins, RNA, and DNA.\\

Overall, these studies in the quantum chemistry of biomolecules confirm the view put here: the quantum dynamics determined by the linear equation (\ref{eq:Schr}) is a locally valid description but does not apply even to complex biomolecules as a whole, much less to living cells or other biological structures.  Their highly non-linear structure and function is enabled by local wave functions everywhere, but not by a single wave function that applies on macro scales. 

\subsection{Quantum circuits and qubits}
A new development is simulation of molecules using superconducting quantum processors 
based in Kitaev-Heisenberg models and highly non-linear quantum circuits (\cite{Tazhigulov}, see eqn.(2) and Figure 2). Does this undermine what has been stated above? Is there a single wave function determining this dynamics?  No, because critical to implementing a quantum computer is the ability to control the state of the qubit.  A quantum algorithm is specified as a set of unitary transformations  $\{U_1, U_2, U_3, ...,\}$  that are implemented by a sequence of Hamiltonians $\{H_1, H_2, H_3, ...,\}$ that are turned on and off one after the other \cite{DiVincenzo}. 
This is done \cite{McKay et al} by a local oscillator  shaped by an Arbitrary Waveform Generator (AWG) which outputs a programmable voltage output $V(t)$   via the classical components shown in Figure 1 of that paper. This controls rotations of the Bloch sphere. Thus the quantum processes involved are controlled in a top-down way by classical components with non-linear branching dynamics, in agreement with this paper.

\subsection{Essentially quantum effects (coherence, entanglement, tunneling) in biology}

Essentially quantum effects in biology are considered in  \cite{Vedral quantum bio} \cite{Lambert quantum bio} \cite{Marais Qu Biol} \cite{McFadden and Al-Khalili}   \cite{Cao quantum biology} \cite{Kim quantum} \cite{Smith avian compass} and \cite{NAS Physics of Life}. Claims in this regard must be treated with caution: there is some very questionable stuff out there, particularly because the brain is warm and wet,  so decoherence is ubiquitous \cite{Tegmark decohere}. The brain is therefore probably not a quantum computer \cite{Tegmark quantum compute}. 

Essentially quantum effects claimed are, 

1. Magnetic field detection by birds has been claimed to be due to sustained quantum coherence and entanglement in the avian compass
\cite{Gauger avian}, but this must be analysed in a realistic manner \cite{Smith avian compass}.

2. Photon to free electron conversion by chlorophyll and rhodopsin molecules %: energy transfer in the photosynthetic antenna
 is  direct evidence that the process involves quantum mechanical coherence \cite{Engel quantum photosynth} \cite{Collini QM Photo} (\cite{NAS Physics of Life}:50-54). Electron transfer in photosynthesis at biologically relevant temperatures depends 
on an interplay of classical and quantum dynamics (\cite{NAS Physics of Life}:52). However this claim must also be examined in the light of realistic biological contexts \cite{Cao quantum biology}.

3. Proton tunneling effects apparently play an important role in enzyme functions \cite{Masgrue quantum tunneling}  (\cite{NAS Physics of Life}:53), which are key in molecular biology.\\

\textbf{The brain} Overall as regards the brain, \cite{Litt et al quantum brain} summarize:
\begin{quote}
	``\textit{We argue that computation via quantum mechanical processes is irrelevant to explaining how brains produce thought, contrary to the ongoing speculations of many theorists. First, quantum effects do not have the temporal properties required for neural information processing. Second, there are substantial physical obstacles to any organic instantiation of quantum computation. Third, there is no psychological evidence that such mental phenomena as consciousness and mathematical thinking require explanation via quantum theory. We conclude that understanding brain function is unlikely to require quantum computation or similar mechanisms},''
\end{quote}
but see  \cite{Fisher qM brain_2} for a useful  survey,  and a claim of entanglement processes in memory. 

Essentially quantum effects plausibly claimed to exist in the brain are always purely local, agreeing with the claim (\S2.3) that there is no wave function for an organism or  brain as a whole. 
 
\subsection{Causal closure revisited}
How do the time-dependent constraints that underlie downward causation to the physics level \cite{Ellis and Kopel} occur? In the context of macromolecular chemistry \cite{Karplus muotiscale}, this occurs via the Born-Oppenheimer approximation discussed previously. The positions $\textbf{R}_i$ of the nuclei in (\ref{eq:Hamiltonian}) are determined by conformational changes of biomolecules as discussed above: 
\begin{equation}\label{eq:positions}
	\textbf{R}_i = \textbf{R}_i(t) \Rightarrow H = H(t) 
\end{equation}
and the evolution determined by (\ref{eq:Schr}) is no longer time reversible; it is determined by the changing biological context, where %the Second Law of Thermodynamics applies and
 higher level dynamics shape outcomes.

 For example, suppose Sue is walking down the street and sees an automobile accident at time $t_1$. This triggers her brain to determine what happened via computational processes  at the emergent brain level \cite{Churchland and Sejnowski}, thereby shaping her immediately successive thoughts (``I must call an ambulance'') based in her character (helpful) and previous knowledge (ambulances bring help).  
Prior knowledge of the state of every particle in her body at any time $t < t_1$ cannot predict this outcome of her brain state at times $t > t_1$, because the accident occurred at an external location: the atoms and electrons in her brain were simply not involved. Causal completeness based in the total details of her initial brain  state is  not possible. Incoming data is handled predictively \cite{Clark} and shapes the pattern of action potential spike chains in her brain to reflect her choice of action for times $t > t_1$ via a process of interlevel causal closure involving such spike chains, which are shaped by rational thought  that alters the positions $\textbf{R}_i(t)$  of nuclei in the voltage gated ion channels in her brain, as well as past experiences that have shaped neural connections (\cite{Drossel 2023}:\S6). Circular causation takes place \cite{Noble et al 2019} that involves all these levels, and indeed also the level of society in which the action is situated, which leads to the existence and functionality of the ambulance.

\section{Conclusion}

%\subsection{The local quantum view vindicated}
Consideration of the effects discussed here will confirm that the local wavefunction view proposed in Section 2 gives a correct description of what is going on.  

 Quantum theory underlies macromolecular chemistry (\S 4.4), but not by use of the Schr\"{o}dinger equation (\ref{eq:Schr}) for a single wave function \cite{Karplus muotiscale}. Even in the case of simple molecules, and certainly macromolecules, one represents electrons as being influenced in a downward way by nuclei that are solved for separately and then used as a classical  background for the electron motions. This is similar to the way electron motions are calculated in solid state physics contexts such as transistors \cite{Drossel and Ellis computers}, with a crystal lattice providing a background for existence of quasi-particles such as phonons 
 that interact with the electrons \cite{Ellis Emerge SS and biology}.  
  Essentially quantum effects occur in very few cases (\S 4.5).
  
    The local wavefunction view put here (\S 2.3) covers all these cases successfully, and is adequate to determine how biology arises from physics, in contrast to the proposal \cite{Carroll (2022)} that there is a single global wave function that will suffice to determine the dynamics of everything.  
There are local wavefunctions everywhere, but no single wavefunction for complex systems, or even for any feedback control loops. There is therefore no wave function for cells, hearts, or  brains as a whole. \textit{Inter alia}, this appears to completely undermine any ``many minds'' worldview based in some version of the Everett interpretation \cite{Zeh} - provided we believe as usual that mental functions are based in brain operations \cite{Kandell}.

\subsection{Domains of validity}
The issue underlying this all is, What is the domain of validity of a theory?  The meta issue is,\begin{quote}
	 \textit{Each theory has a restricted domain of application}.
\end{quote} Newtonian physics is very good within its domain of application;
so is Galilean gravity, Newton's theory of gravity, Maxwell's theory of electromagnetism, Einstein's theory of gravitation, and so on. But each has its limits, as is well known; other theories take over outside that domain.

The question not asked often at present is,  setting aside the issue of quantum gravity (where it is indeed often asked), 
\begin{quote}
	\textit{What is the domain of application of Quantum Theory in a specific context?}
\end{quote}
Nowadays it is often taken to be universally applicable at all scales, for example to cats, brains, and indeed the universe as a whole. 
But it is no exception to the general rule just stated: its domain of application is limited too. For example, it does not apply as such to heat baths \cite{Drossel 10 reasons} or feedback control (this paper, \S2.2, \S2.3), cats, or brains (\S2.4). 

Rather quantum theory is locally applicable everywhere at all times in domains small enough  that the essential dynamics is linear \cite{Ellis (2012) QM}. Local wavefunctions allow this, as explained in this paper. Adaptive modular hierarchical structures then enable complex emergence out of these simple local dynamics (\S3); and those structures are classical. There is no single wavefunction that underlies their existence,
 %\textit{Wave functions and levels} 
 particularly because they are multiscale \cite{multiscale}. %Suppose a single wavefunction $|\psi_1\rangle$ were to apply at level $L1$ for a living entity as a whole. %Is it then plausible that a single wave function $|\psi_2\rangle$,  determined by some kind of coarse graining from level $L1$ to level $L2$, would apply at a higher level $L2$? This is problematic because c
 %Coarse-graining does not generically preserve linearity \cite{Ellis (2012) QM}, after all dissipative processes emerge. %But suppose that that was, against all probabilities, true. Then $|\psi_1\rangle$ and $|\psi_2\rangle$ would be different wave functions for the system (in the case of the kinetic theory of gases, $|\psi_1\rangle$ would characterise particle properties such as position and momentum, while $|\psi_2\rangle$  would characterise macro properties such as pressure and density). It would not be described by any single wave function, because 
 %Emergent effective dynamics is different at each emergent level \cite{Ellis Emerge SS and biology}: indeed, that is how they are identified as being emergent levels. A single wave function cannot describe a real world emergent hierarchical structure. 

This is true not only in biology but also for example in engineering contexts such as digital computers \cite{Drossel and Ellis computers}, and so in aircraft, cellphones and so on. It also applies to apparatus used in quantum physics experiments: which can therefore be thought of as classical. %And if a hidden variable theory were true, it would arguably apply equally in this case. 

\subsection{Testing domains of validity} 
The key question then is, \begin{quote}
	\textit{How does one test the domain of validity of quantum theory in a specific context}?
\end{quote}
\textbf{Constructively}, one does so by engineering situations where wave functions are not local, as in the experiments on entangled photons that won the 2022 Noble Prize in Physics for Aspect, Clauser, and Zeilinger. In this case violations of Bell inequalities, or quantum teleportation,  established that there was a macroscopic domain of validity for a wave function. But those were not situations that occur naturally: they were carefully shaped precisely so as to give this result. 

The claims of this paper are not negated by a remarkable experiment \cite{tardigrade} where a tardigrade - a microscopic multicellular organism
known to tolerate extreme physiochemical conditions via a latent state of life known as cryptobiosis - was entangled with superconducting qubits. The experiment involved putting the tardigrade in an environment where linearity could occur, namely a temperature of sub $10$ mK temperatures and pressure of $6 \times 10^{−6}$ mba, hence preventing any significant interaction with heat baths. No metabolism or feedback loops were operational in these conditions: the organism was in a state of suspended animation, effectively being a single frozen crystal.  Because of the extraordinary nature of the organism, this did not destroy 
its ability to function biologically when reverted to ordinary conditions for life. But it was never both entangled and functioning biologically at the same time. This situation is of course highly artificial, and cannot occur naturally, \textit{inter alia} because the entire Universe is pervaded by  primordial black body radiation at a temperature of $2.73 K$ ($\gg 10 mK$) at the present time  \cite{HE}; the requisite temperature can only be created artificially. From the viewpoint of this paper, the experiment demonstrates how the domains of  validity of wavefunctions for a specific entity  can    vary with time, depending on context. \\

\textbf{Negatively,} unless you believe standard quantum theory to be committed to macrorealism (agreed on all sides to be an unfortunate name),  one can examine cases where standard quantum theory can be shown to be invalid, as in the case of the Leggett-Garg inequalities \cite{Leggett and Garg} \cite{Timpson}. Experiments based on these inequalities can be carried out to rule out quantum macrorealism, as for example in \cite{Kne et al legget garg}. This establishes  the principle of tests for macrorealism in larger system. One can also establish no-go theorems pertaining to the existence of long range entanglement by giving upper bounds to quantum correlations \cite{Kuwahara and Saito}.

There is a community of quantum physicists and philosophers looking at these issues, see for example \cite{Oxford questions},  \cite{Ares Briggs}.  I will not attempt an in-depth analysis here; %I simply point out that 
many of the issues discussed there either directly relate to the proposals of the present paper, or can be adapted to do so. Perhaps the main thing missing in those papers is an interaction with the literature on quantum chemistry. This paper fills that gap to some extent.

\subsection{Irreversibility and the direction of time} 
The Schr\"{o}dinger equation (\ref{eq:Schr}) and Hamiltonian (\ref{eq:hamiltonian}) are time symmetric, so where does the arrow of time associated with quantum physics outcomes come from? How does an essentially time symmetric quantum theory produce the arrow of time in biology?

The basic time asymmetry that occurs in the physical universe is the expansion of the Universe itself \cite{Ellis Time}. This determines a global \textit{Direction of Time} that points from the start of the Universe to the present day. Together with special conditions in the early universe \cite{Layzer} \cite{Albert time}, this reaches down to affect all the local \textit{arrows of time} (electrodynamic, thermodynamic, wave,  diffusion, fracture,  biological, mental, and quantum) \cite{Ellis time 2014} \cite{Ellis Time} \cite{Drossel and Ellis time}. The quantum arrow of time  arises firstly because  realistic quantum collapse processes interact with heat baths  \cite{Drossel and Ellis quantum}, and so are time asymmetric \cite{Drossel and Ellis time}. Secondly, local microbiology functioning affected by the Second Law of Thermodynamics \cite{Eddington A S (1927)} reaches down to alter positions of molecules as in (\ref{eq:positions}), and hence changes the Hamiltonian (\ref{eq:hamiltonian}) in a time dependent way. 
Hence the evolution determined by (\ref{eq:Schr}) together with (\ref{eq:Hamiltonian}),  (\ref{eq:positions}), as well as the wave function collapse process and associated loss of of information, are both time asymmetric \cite{Drossel 2023}.

We can of course measure the passing of time via suitable clocks, and an interesting aspect is the feedback that occurs as regards the Second Law:  any system which measures the passage of time dissipates entropy at a rate that is proportional to the accuracy of the timekeeping \cite{Pearson et al}. This is an emergent classical relation with the correct direction of time.\\

\textbf{Acknowledgements}: I thank Andrew Briggs,  Howard Wiseman, and Ruth Kastner for helpful comments, Martin Ciupa for a helpful reference, and particularly Jeremy Butterfield for a careful reading of the paper leading to comments resulting in many improvements.  

I thank  the University of Cape Town Research Committee for financial support.

\small {Version: 2023/02/24}. 
\end{document}